\documentclass[a4paper,14]{article}
\usepackage[english]{babel}
\usepackage{amsmath,amsfonts,amstext}
\usepackage{graphicx}
\usepackage[latin1]{inputenc}
\usepackage{multicol}
\usepackage{setspace}
\usepackage{bm}
\usepackage[numbers]{natbib}
\newtheorem{Def}{Definition}
\newtheorem{Prop}{Proposition}
\newcommand\indep{\protect\mathpalette{\protect\independenT}{\perp}}
\def\independenT#1#2{\mathrel{\rlap{$#1#2$}\mkern2mu{#1#2}}}

\newcommand{\y}{\mathcal{Y}}

\newcommand\PathFig{}

\title{An EM  algorithm for estimation in the  Mixture Transition Distribution
  model}
\author{Sophie L\`ebre$^{\ast}$\thanks{$^{\ast}$ To whom correspondence should 
  be addressed.}, Pierre-Yves Bourguignon \\
  Laboratoire Statistique et G\'enome,\\
  UMR 8071 Universit\'e Evry Val d'Essonne/CNRS UMR8071/INRA 1152\\ 
  523, place des Terrasses de l'Agora, 91000 Evry, France.} 

\date{}

\begin{document}

\maketitle
 
\markboth{An EM algorithm for estimation in the MTD model}{S. L\`ebre, P.-Y. 
 Bourguignon}

\begin{abstract}
The Mixture Transition Distribution (MTD) model was introduced by Raftery to face the need for parsimony in the modeling of high-order Markov chains in discrete time. The particularity of this model comes from the fact that the effect of each lag upon the present is considered separately and additively, so that the number of parameters required is drastically reduced. 
However, the efficiency for the MTD parameter estimations proposed up to date still remains problematic on account of the large number of constraints on the parameters. In this paper, an iterative procedure, commonly known as Expectation-Maximization (EM) algorithm, is developed cooperating with the
principle of Maximum Likelihood Estimation (MLE) to estimate the MTD parameters. Some applications of modeling MTD show the proposed EM algorithm is easier to be used than the algorithm developed by Berchtold. Moreover, the EM Estimations of parameters for high-order MTD models led on DNA sequences outperform the corresponding fully parametrized Markov chain in terms of Bayesian Information Criterion.

A software implementation of our algorithm is available in the library seq$++$
at \texttt{http://stat.genopole.cnrs.fr/seqpp}.

\textit{keywords:} Markov chain; mixture transition distribution (MTD);
Parsimony; Maximum likelihood; EM algorithm;

\end{abstract}

\large

\section{Introduction}\label{section:intro}

While  providing  a  useful  framework for  discrete-time  sequence  modeling,
higher-order Markov chains suffer from the exponential growth of the parameter
space dimension with  respect to the order of the model,  which results in the
inefficiency of the  parameters'estimations when a limited  amount of data is
available.  This  fact motivates the  developments of approximate  versions of
higher-order Markov chains, such  as the Mixture Transition Distribution (MTD)
model    \cite{raftery1,br02}    and    variable    length    Markov    chains
\cite{buhlman99}. 
Thanks to a simple structure, where  each lag contributes to the prediction of
the current  letter in  a separate  and additive way,  the dimension  of model
parameter  space grows  only linearly  with respect  to the  order of  the MTD
model.

Nevertheless, Maximum Likelihood Estimation (MLE)  in the MTD model is subject
to such constraints that analytical  solutions are beyond the reach of present
methods. One has thus to retort to numerical optimization procedures. The most
powerful method proposed to this day is due to Berchtold \cite{berchtold}, and
relies on an ad-hoc optimization method.  In this paper, we propose to fit the
MTD model into  the general framework of hidden variable  models, and derive a
version of the classical EM algorithm for the estimations of its parameters.

In this  first section, we define the  MTD model and recall  its main features
and some of its variants. Parametrization of the model is discussed in section
2,  where we  establish that  under  the most  general definition,  it is  not
identifiable.  Then  we shed  light  on  an  identifiable set  of  parameters. 
Derivations of the  update formulas involved by the  EM algorithm are detailed
in  section  3. We  finally  illustrate our  method  by  some applications  to
biological sequence modeling.

\paragraph{Need for parsimony}

Markov  models are  pertinent to  analyze $m$-letter  words' composition  of a
sequence  of  random  variables  \cite{fichan87,krogh99}.   Nevertheless,  the
length  $m$ of  the words  the model  accounts  for has  to be  chosen by  the
statistician.  On the one hand, a  high order is always preferred since it can
capture strictly more  information.  On the other hand,  since the parameter's
dimension increases exponentially fast with respect to the order of the model,
higher order models  cannot be accurately estimated. Thus,  a trade-off has to
be drawn to optimize the amount of information extracted from the data.

%%%%%%%%%%%%%%%%%%%%%%%%%%%%%%%%%%%%%%%%%%%%%%%%%%%%%%%%%%%%%%%%%%%%%%%%%%%%
\begin{figure}
\centering%
\includegraphics[width=.6\textwidth]{\PathFig 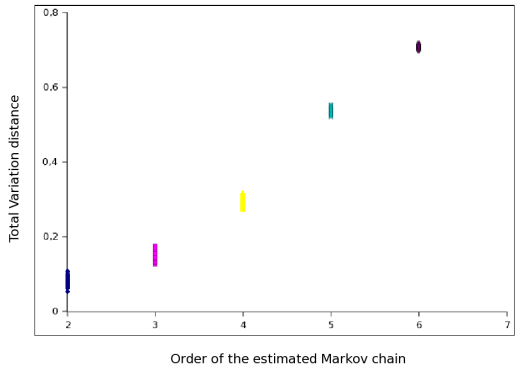}
\caption{Total  variation   distance  between  distributions   estimated  from
  randomly generated sequences and the generating distribution. The generating
  model  is   of  order  5,  and   the  random  sequences   are  5000  letters
  long.}\label{fig:dvt}
\end{figure}
%%%%%%%%%%%%%%%%%%%%%%%%%%%%%%%%%%%%%%%%%%%%%%%%%%%%%%%%%%%%%%%%%%%%%%%%%%%%

We  illustrate this  phenomenon by  running a  simple experiment:  by  using a
randomly  chosen Markov chain  transition matrix  of order  5, we  sample 1000
sequences of length 5000. Each of them is then used to estimate a Markov model
transition matrix of  order varying from 2 to 6. For  each of these estimates,
we have  plotted the total variation  distance with respect  to the generating
model    (see    Figure    \ref{fig:dvt}),    computed   as    the    quantity
$D_{VT}(P,Q)=\sum_{x\in \mathcal{Y}^n}|P(x) - Q(x)|$ for distributions $P$ and
$Q$.  It  turns out that  the optimal estimation  in terms of  total variation
distance between genuine and estimated  distributions is obtained with a model
of order 2 whereas the generating model is of order 5.

Mixture Transition Distributions  aim at providing a model  accounting for the
number  of occurrences  of $m$-letter  words, while  avoiding  the exponential
increase with  respect to $m$ of  the full Markov  model parameter's dimension
(See Table \ref{tab:nbPar} for a comparison of the models' dimensions).

\paragraph{MTD modeling}
Let $\bm{Y}=(Y_1,\dots,Y_n)$  be a sequence of random  variables taking values
in  the  finite set  $\mathcal{Y}=\lbrace  1,\dots,  q  \rbrace$. We  use  the
notation, $$
\bm{Y}_{t_1}^{t_2}=(Y_{t_1}, Y_{t_1+1},\dots,Y_{t_2}) $$
\noindent  to   refer  to  the  subsequence  of   the  $t_2-t_1+1$  successive
variables. In the whole paper, vectors and matrices are denoted by bold letters.

\begin{Def} The random  sequence $\bm{Y}$ is said to be  an $m^{th}$ order MTD
  sequence if
\begin{eqnarray}\label{defg}
  \forall t>m, \forall y_1,\dots, y_t \in \mathcal{Y}, \ \ \mathbb{P}(Y_t=y_t 
  \vert \bm{Y}_1^{t-1}=\bm{y}_1^{t-1})
  &=&\sum_{g=1}^{m}\varphi_g \ \mathbb{P}(Y_t=y_t\vert Y_{t-g}=y_{t-g})\nonumber \\
  &=&\sum_{g=1}^{m}\varphi_g\bm{\pi}_g(y_{t-g},y_t).
\end{eqnarray}
\noindent  where   the  vector  $\bm{\varphi}=(\varphi_1,\dots,\varphi_m)$  is
subject to the constraints:
\begin{eqnarray}
  \forall g \in \lbrace 1,\dots,m\rbrace , \ \varphi_g  \geq 0,\label{phiPos}\\
  \sum_{g=1}^{m}\varphi_g=1.\label{phiSum1}
\end{eqnarray}
\noindent  and  the  matrices $\lbrace  \bm{\pi_g}=\left[\mathbb{P}(Y_t=j\vert
  Y_{t-g}=i)\right]_{i,j \in  \y}; \ 1\leq g  \leq m\rbrace$ are  $q \times q$
stochastic matrices.
\end{Def}

A $m$th-order MTD model is thus defined by a vector parameter,
\begin{equation*}
  \bm{\theta}= \left( \varphi_g,\left( \pi_g(i,j)\right)_{ i,j \in \y}\right)_{1\leq g \leq m}
\end{equation*}
\noindent which belongs to the space
\begin{eqnarray*}
\Theta&=&\left\lbrace \bm{\theta}; \ \forall \ 1 \leq g \leq m, \, 0
  \leq \varphi_g \leq 1\, ; \, \sum_{g=1}^m \varphi_g =1\, ; \right.\\
&&\left.  \forall i,j  \in  \y, \,  0\leq  \pi_g(i,j) \leq  1\, \text{and}  \,
  \sum_{j \in \y} \pi_g(i,j)=1 \right\rbrace. 
\end{eqnarray*}

It is  obvious from the first  equality in equation (\ref{defg})  that the MTD
model fulfills  the Markov property. Thus,  MTD models are  Markov models with
the particularity that each lag $Y_{t-1},Y_{t-2},\dots$ contributes additively
to  the distribution  of the  random  variable $Y_t$.   Berchtold and  Raftery
\cite{br02}  published  a complete  review  of  the  MTD model.   They  recall
theoretical  results  on  the  limiting  behavior  of the  model  and  on  its
auto-correlation  structure.  Details  are given  about several  extensions of
this model, such as infinite-lag  models, or infinite countable and continuous
state space.

We have to  point out that Raftery \cite{raftery1}  defined the original model
with   the  same   transition  matrix   $\bm{\pi}$  for   each   lag  $\lbrace
Y_{t-g}\rbrace_{g=1,\dots,m}$.  In the  sequel, we refer to this  model as the
\textit{single}  matrix   MTD  model.   Later,   Berchtold  \cite{berchtold95}
introduced  a more  general  definition of  the  MTD models  as  a mixture  of
transitions  from  different  subsets  of lagged  variables  $\{Y_{t-m},\dots,
Y_{t-1}\}$  to  the present  one  $Y_t$,  eventually  discarding some  of  the
dependencies.  In  this paper,  we focus on  a slightly more  restricted model
having a specific  but same order transition matrix  $\bm{\pi}_g$ for each lag
$Y_{t-g}$. We denote by MTD$_l$ the MTD model which has a $l$-order transition
matrix for each  lag (Definition \ref{def:MTDl}).  From now  on, the MTD model
defined by (\ref{defg}) is denoted accordingly by MTD$_1$.

\begin{Def}\label{def:MTDl}  The  random  sequence  $\bm{Y}$  is  a  $m$-order
  MTD$_l$ sequence if,  for all $l,m \in \mathbb{N}$ such  that $l<m$, and all
  $\bm{y}_1^t \in \mathcal{Y}^t$ :
\begin{eqnarray*}
  \mathbb{P}(Y_t=y_t \vert \bm{Y}_1^{t-1}=\bm{y}_1^{t-1})&=&\mathbb{P}(Y_t=i_t \vert \bm{Y}^{t-1}_{t-m}=\bm{y}^{t-1}_{t-m})\\
  &=&\sum_{g=1}^{m-l+1}\varphi_g \ \mathbb{P}(Y_t=y_t\vert \bm{Y}_{t-g-l+1}^{t-g}=\bm{y}_{t-g-l+1}^{t-g})\\
  &=&\sum_{g=1}^{m-l+1}\varphi_g \ \bm{\pi}_g(\bm{y}_{t-g-l+1}^{t-g},y_t).
\end{eqnarray*}
holds, where $\bm{\pi}_g$ is a $q^l \times q$ transition matrix.
\end{Def}

\paragraph{Trade-off between dimension and maximal likelihood}

Even though MTD  models involve a restricted amount  of parameters compared to
Markov chains, increasing the order $l$  of the model may result in efficiency
of the MLE decreased. The quality of  the trade-off
between  goodness-of-fit and  generalization  error a  model  achieves can  be
assessed  against classical  model selection  criteria, such  as  the Bayesian
Information       Criterion      (see      illustrations       in      section
\ref{subsection:estimDNA}).

However,  computing the BIC  requires the  knowledge of  the dimension  of the
model. This  dimension is usually computed  as the dimension  of the parameter
space for a bijective parametrization. In the specific case of the MTD models,
the original single-matrix  model is parametrized in a  bijective way, whereas
its  generalized version  with specific  transition matrices  for each  lag is
over-parametrized: in  appendix \ref{appendix:2eq} is given an  example of two
distinct values of the parameters $(\bm{\varphi},\bm{\pi})$, which both define
the same MTD$_1$ distribution.  The dimension  of the model is thus lower than
the  dimension  of  the parameter  space,  and  computing  the BIC  using  the
parameter  space dimension  would over-penalize  the models.  A  tighter upper
bound  of  the   dimension  of  the  MTD$_l$  model   is  derived  in  section
\ref{section:parms}, a bound which is used later to compute the BIC.

\paragraph{The question of estimation}

As  a counterpart  for their  parsimony, MTD  parameters are  difficult  to be
estimated due  to the constraints  that the transition  probabilities $\lbrace
\mathbb{P}(i_m\dots i_1;i_0);$ $ i_m,\dots,  i_0 \in \y\rbrace$ have to comply
to.  There  is  indeed no  analytical  solution  to  the maximization  of  the
log-likelihood  $L_y(\bm{\theta})=\mathbb{P}_{\bm{\theta}}(\bm{Y}=\bm{y})$  of
the  MTD  models under  the  constraints  the  vector $\bm{\varphi}$  and  the
stochastic  matrices  $\bm{\pi}_g$ have  to  fulfill.   For  a given  sequence
$\bm{y}=y_1,\dots,y_n$ of length $n$, we  recall that the loglikelihood of the
sequence $\bm{y}$ under the MTD$_1$ model writes
\begin{eqnarray*}
L_y(\theta)&=&\log \mathbb{P}_{\theta}(\bm{Y}_1^n=\bm{y}_1^n)\\
&=&   \log   \left\lbrace   \mathbb{P}(\bm{Y}_1^m=\bm{y}_1^m)   \prod_{t=m+1}^n   \left(
    \sum_{g=1}^m \varphi_g \bm{\pi}_g(y_{t-g},y_t)\right) \right\rbrace. 
\end{eqnarray*}

The estimation of  the original single matrix MTD model  already aroused a lot
of interest.  Although any distribution from this model is defined by a unique
parameter  $\bm{\theta}$,  the  maximum  likelihood can  not  be  analytically
determined.  Li  and Kwok \cite{likwok} propose an  interesting alternative to
the maximum  likelihood with a  minimum chi-square method.  Nevertheless, they
carry out estimations by using a non-linear optimization algorithm that is not
explicitly   described.     Raftery   and   Tavar{\'e}    \cite{rt94}   obtain
approximations  of both  maximum likelihood  and minimum  chi-square estimates
with numerical procedures from the  NAG library which is not freely available. 
They also  show that the  MTD model can  be estimated using  GLIM (Generalized
Linear Interactive Modeling) in the specific case where the state space's size
$q$  equals  2.   Finally,  Berchtold  \cite{berchtold} developed  an  ad  hoc
iterative  method implementing  a constrained  gradient  descent optimization. 
This algorithm is  based on the assumption that  the vector $\bm{\varphi}$ and
each row of the matrix $\bm{\pi}$ are independent. It consists in successively
updating each of  these vectors constrained to have a  sum of components equal
to 1 as follows.

\paragraph{Berchtold's Algorithm}
\textit{
\begin{itemize}
\item  Compute partial  derivatives of  the log  likelihood according  to each
  element of the vector,
\item choose a value $\delta$ in $[0,1]$,
\item add  $\delta$ to the the component with the  largest derivative,
  and  subtract $\delta$  from  the one with the  smallest derivative.
\end{itemize}
}

This algorithm  has been shown  to perform at  least better than  the previous
methods, and it  can be extended to  the case of the MTD$_l$  models.  In this
latter case, it estimates  \textit{one} of the parameter vectors $\left\lbrace
  (\varphi_g,\bm{\pi}_g);  1  \leq  g  \leq m\right  \rbrace$  describing  the
maximum-likelihood  MTD   distribution.   Nevertheless,  the   choice  of  the
\textit{alteration  parameter} $\delta$  remains an  issue of  the  method. An
in-depth discussion  of the strategy  used to update the  alteration parameter
$\delta$ can be found in \cite{berchtold}.

We propose  to approximate  the maximum likelihood  estimate of the  MTD model
$\left\lbrace  \hat{\mathbb{P}}_{ML}(i_m\dots  i_1;i_0);  i_m,\dots,  i_0  \in
  \y\right \rbrace$  by coming down to  a better known  problem: estimation of
incomplete data with an Expectation-Maximization (EM) algorithm \cite{DLR}. We
introduce a simple estimation  method which allows to approximate \textit{one}
parameter  vector $\theta=\lbrace (\varphi_g,\bm{\pi}_g);  \ 1  \leq g  \leq m
\rbrace$ maximizing the log-likelihood.

\section{Upper bound of the MTD model dimension}\label{section:parms}

The MTD$_1$ model is over-parametrized.  We provide an example of two distinct
parameter  values $(\bm{\varphi},\bm{\pi})$  defining the  same $2^{nd}$-order
MTD$_1$  model in  appendix \ref{appendix:2eq}.   Moreover, we  propose  a new
parameter  set whose  dimension is  lower. It  stems from  the straightforward
remark that the $m$th-order MTD$_1$ model satisfies the following proposition:

\begin{Prop}\label{diffg}
\mbox{Transition probabilities of a $m$th-order MTD${}_1$ model satisfy:}
\begin{multline}
  \forall i_m,...,i_g,...,i_0,i_g' \in \mathcal{Y},\\
  \mathbb{P}(i_m...i_g...i_1;i_0)-\mathbb{P}(i_m...i_g'...i_1;i_0)=\varphi_g\left[\bm{\pi}_g(i_g,i_0)-\bm{\pi}_g(i_g',i_0)\right].
\label{eq:diffTransProb}
\end{multline}
\end{Prop}
This simply means that the left-hand side of equation (\ref{eq:diffTransProb})
only depends on the parameter components associated to lag $g$.

Consider    a    given   distribution    from    MTD${}_1$   with    parameter
$(\varphi_g,\bm{\pi}_g)_{1\le g \le  m}$, and let $u$ be  an arbitrary element
of $\mathcal{Y}$. Each  transition probability $\mathbb{P}(i_m...i_1;i_0)$ may
be written :
\begin{equation}\label{probam}
\mathbb{P}(i_m...i_1;i_0)= \sum_{g=1}^m  \varphi_g \left[ \bm{\pi}_g(i_g,i_0)-
  \bm{\pi}_g(u,i_0) \right] 
+ \sum_{g=1}^m \varphi_g \bm{\pi}_g(u,i_0).
\end{equation}
From  Proposition \ref{diffg},  it follows  that each  term of  the  first sum
$\varphi_g \left[  \bm{\pi}_g(i_g,i_0)- \bm{\pi}_g(u,i_0) \right]$  equals the
difference                           of                          probabilities
$\mathbb{P}(u...ui_gu...u;i_0)-\mathbb{P}(u...u;i_0)$.   The   second  sum  is
trivially the  transition probability from  the $m$-letter word $u\dots  u$ to
$i_0$.

Let  us denote  the  transition  probabilities from  $m$-letter  words to  the
letter~$j$,  restricting to words  differing from  $u\dots u$  by at  most one
letter :
\begin{equation}
\label{eq:newParam}
p_u(g \ ;i,j):=\mathbb{P}(u...uiu...u;j),
\end{equation}
\noindent where $u...uiu...u$ is the  $m$-letter word whose letter in position
$g$ (from  right to left) is  $i$.  The quantities  in (\ref{eq:newParam}) are
sufficient to describe the model, as stated in the following proposition.

\begin{Prop}\label{prop:parms}
\mbox{The transition probabilities of a $m$th-order MTD$_1$ model satisfy:}
\begin{multline*}
  \forall u \in \mathcal{Y}, \ \forall i_m,...,i_g,...,i_0 \in \mathcal{Y},\\
  $$\mathbb{P}(i_m,...,i_1;i_0)= \sum_{g=1}^m  \left [ p_u(g\ ;i_g,i_0)  \ - \
    \frac{m-1}{m} \ p_u(i_0)\right].$$
\end{multline*}
where $p_u(j)$ denotes the value of $p_u(g;u,j)$, whatever the value of $g$.
\end{Prop}

For any arbitrary $u$ element  of $\mathcal{Y}$, a MTD$_1$ distribution can be
parametrized by a vector $\theta_u$ from the $(q-1)[1+m(q-1)]$-dimensional set
$\bar{\Theta}_u$,
\begin{multline}\label{theta_u}
  \bar{\Theta}_u=\left   \lbrace   ((p_u(g;i,j))_{1\le   g\le   m,   i,j   \in
      \mathcal{Y}} \text{ such that } \forall g \in \lbrace 1,\dots,m\rbrace,
    \forall    i   \in\mathcal{Y},   \phantom{\sum_{j\in\mathcal{Y}}}\right.\\
  \left.  \sum_{j\in\mathcal{Y}}   p_u(g;i,j)  =   1  \text{  and   }  \forall
    g,g^{\prime}     \in\{1,\dots,m\},    p_u(g;u,j)=p_u(g^{\prime};u,j)\right
  \rbrace
\end{multline}
Note  that  not  all  $\bm{\theta}_u$  in $\bar{\Theta}_u$  define  a  MTD$_1$
distribution: the sum ${\sum_{g=1}^m p_u(g;i_g,i_0) - \frac{m-1}{m} p_u(i_0)}$
may indeed  fall outside the  interval $[0,1]$. For  this reason, we  can only
claim that some subset $\Theta_u$ of $\bar{\Theta}_u$ is a parameter space for
the   MTD$_1$   model.    However,   as   the  components   of   a   parameter
$\theta_u\in\Theta_u$  are transition  probabilities, two  different parameter
values can not define the same MTD distribution.  The mapping of $\Theta_u$ on
the  MTD$_1$  model is  thus  bijective, which  results  in  the dimension  of
$\bar{\Theta}_u$ being an upper bound of the dimension of the MTD model.

Whereas the original definition of  the MTD$_1$ model (\ref{defg}) involves an
$m-1+mq(q-1)$-dimensional parameter  set, this  new parametrization lies  in a
smaller dimensional space, dropping $q(m-1)$ parameters.

%%%%%%%%%%%%%%%%%%%%%%%%%%%%%%%%%%%%%%%%%%%%%%%%%%%%%%%%%%%%%%%%%%%%%%%%%%%%
\begin{table}
\begin{center}
  \caption{\textbf{Number of independent  parameters required to describe full
      Markov and  MTD$_l$ models (state  space size: $q=4$).}  Except  for the
    single  matrix MTD model,  MTD models  originally defined  with parameters
    $(\varphi,\bm{\pi})$     are    over    parametrized:     the    parameter
    $\bm{\theta}_u^l$, introduced in section \ref{section:parms}, requires far
    less independent  parameters. Note that  the $1^{st}$ order  MTD$_1$ model
    (resp. $2^{nd}$ order  MTD$_2$ model) is equivalent to  the $1^{st}$ order
    (resp. $2^{nd}$ order) full Markov model.}
\label{tab:nbPar}
\begin{tabular}{|c|c|c|c|c|c|}
  \hline
  &Full&\multicolumn{2}{|c|}{\mbox{MTD$_1$}}&\multicolumn{2}{|c|}{\mbox{MTD$_2$}}\\
  \cline{3-6}
  $\mbox{Order    }   m$&$\mbox{Markov}$&$    \vert(\bm{\varphi}$,$\bm{\pi})    \vert   $&
  \phantom{\LARGE{1}} $ \vert \bm{\theta}_u^1 \vert $ \phantom{\LARGE{1}} &$ \vert
  (\bm{\varphi}$,$\bm{\pi}) \vert$ &\phantom{\LARGE{1}} $ \vert\bm{\theta}_u^2 \vert$ \phantom{\LARGE{1}} \\
  \hline
  1&12&12&12&&\\
  2&48&25&21&48&48\\
  3&192&38&30&97&84\\
  4&768&51&39&146&120\\
  5&3 072&64&48&195&156\\
  \hline
\end{tabular}
\end{center}
\end{table}
%%%%%%%%%%%%%%%%%%%%%%%%%%%%%%%%%%%%%%%%%%%%%%%%%%%%%%%%%%%%%%%%%%%%%%%%%%%%

Equivalent  parametrization can  be set  for  MTD models  having higher  order
transition matrix  for each lag.   For any $l\geq  1$, a MTD$_l$ model  can be
described  by a  vector  composed of  the  transition probabilities  $p_u^l(g;
i_l...i_1,j)  = \mathbb{P}(u...ui_l...i_1u...u;j)$  for  all $l$-letter  words
$i_l...i_1$.  Denoting by $\Theta_u^l$  the corresponding parameter space, its
dimension $\vert  \Theta_u^l \vert=  \sum_{k=2}^l [ q^{k-2}  (q-1)^3(m-k+1)] +
(1+m(q-1))(q-1)$  is  again  much   smaller  than  the  number  of  parameters
originally  required  to describe  the  MTD$_l$  model (see  \cite{adeline05},
section  2.2, for  the  counting  details).  A  comparison  of the  dimensions
according to  both parametrizations appears in Table  \ref{tab:nbPar}. We will
now make use  of the upper bound $\vert \bm{\theta}_u^l\vert  $ of the model's
dimension  to  penalize  the  likelihood  in  the  assessment  of  MTD  models
goodness-of-fit (see section \ref{subsection:estimDNA}).

\section{Estimation}

In this section,  we expose an EM algorithm for the  estimation of MTD models. 
Firstly, this procedure allows to maximize the likelihood without assuming the
independence  of parameters  $\varphi$ and  $\pi$ and  offers  the convergence
properties of an  EM algorithm. Secondly, from a technical  point of view, the
EM algorithm does not require any  trick to fulfill the constraints holding on
the  ($\varphi$,$\pi$) parameters  as Berchtold's  algorithm does.   We expose
here our estimation method of the MTD$_1$ model (\ref{defg}) having a specific
$1^{st}$  order transition  matrix for  each lag.   The method  can  easily be
adapted  for  single matrix  MTD  models  as well  as  for  MTD models  having
different types  of transition matrix  for each lag.  Detailed  derivations of
the  formulas for identical  matrix MTD  and MTD$_l$  models are  presented in
appendix~\ref{appendix:mtdl}.

To      estimate      the      transition     probabilities      $\left\lbrace
  \mathbb{P}(i_m....i_1;i_0);  i_m,...,  i_0  \in  \y  \right  \rbrace$  of  a
$m$th-order  MTD$_1$  model,  we   propose  to  compute  an  approximation  of
\textit{one} set of  parameters $\bm{\theta}= (\varphi_g,\bm{\pi}_g)_{1 \leq g
  \leq m}$ which maximizes the likelihood.

\subsection{Introduction of a hidden process}
Our approach lies on a  particular interpretation of the model. The definition
of the  MTD$_1$ model (\ref{defg})  is equivalent to  a mixture of  $m$ hidden
models where the  random variable $Y_t$ is predicted by one  of the $m$ Markov
chains $\bm{\pi}_g$  with the corresponding  probability $\varphi_g$.  Indeed,
the coefficients $(\varphi_g)_{g=1,..,m}$ define  a probability measure on the
finite  set  $\lbrace  1,...,m\rbrace$  since  they  satisfy  the  constraints
(\ref{phiPos}) and (\ref{phiSum1}).

From  now  on,  we consider  a  hidden  state  process  $S_1, ...,  S_n$  that
determines  the way  according to  which the  prediction is  carried  out. The
hidden state variables $\lbrace S_t \rbrace $, taking values in the finite set
$\mathcal{S}=\lbrace1,   ...,m\rbrace$,   are   independent  and   identically
distributed, with distribution
\begin{equation*}
\forall t\leq n, \forall g \in \mathcal{S},\ \ \mathbb{P}(S_t=g)=\varphi_g.
\end{equation*}

\begin{figure}
\center
\includegraphics[width=0.7\textwidth]{\PathFig 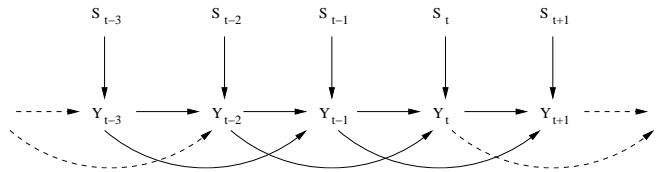}
\caption{DAG dependency structure of a $2^{nd}$ order MTD$_1$ model.} 
\label{fig:dependance}
\end{figure}

The MTD$_1$  model is then  defined as a  hidden variable model.  The observed
variable  $Y_t$ depends  on the  current  hidden state  $S_t$ and  on the  $m$
previous  variables $Y_{t-1},...,Y_{t-m}$.  This dependency  structure  of the
model  is   represented  as   a  Directed  Acyclic   Graph  (DAG)   in  Figure
\ref{fig:dependance}.  The hidden  value at  one position  indicates  which of
those previous  variables of transition  matrices are to  be used to  draw the
current letter: conditional on the state $S_t$, the random variable $Y_t$ only
depends on the variable $Y_{t-S_{t}}$:
\begin{equation*}
\forall t>m, \forall g \in \mathcal{S},\ \ 
\mathbb{P}(Y_t=y_t \vert Y_{t-m}^{t-1}=\bm{y}_{t-m}^{t-1}, S_t=g)=\bm{\pi}_g(y_{t-g},y_t).
\end{equation*}

%%%%%%%%%%%%%%%%%%%%%%%%%%%%%%%%%%%%%%%%%%%%%%%%%%%%%%%%%%%%%%%%%%%%%%%%%%%%
%%%%%%%%%%%%%%%%%%%%%%%%%%%%%%%%%%%%%%%%%%%%%%%%%%%%%%%%%%%%%%%%%%%%%%%%%%%%

So we  carry out estimation in the  MTD$_1$ models as estimation  in a mixture
model where  the components  of the  mixture are $m$  Markov chains,  each one
predicting the variable $Y_t$ from one of the $m$ previous variables.

\subsection{EM algorithm}

By  considering a  hidden  variables model,  we  want to  compute the  maximum
likelihood  estimate from  incomplete  data. The  EM  algorithm introduced  by
Dempster et al. \cite{DLR} is a  very classical framework for achieving such a
task. It has proved to be particularly efficient at estimating various classes
of hidden variable models. We make it entirely explicit in the  case
of the MTD models.

The  purpose  of  the EM  algorithm  is  to  approximate  the maximum  of  the
log-likelihood     of    the     incomplete     data    $L_y(\bm{\theta})=\log
\mathbb{P}_{\bm{\theta}}(Y=y)$  over   $\bm{\theta}  \in  \Theta$   using  the
relationship    $$
\forall    \bm{\theta},   \bm{\theta}'    \in    \Theta,
L_{\bm{y}}(\bm{\theta})=Q(\bm{\theta}  \vert \bm{\theta}')-H(\bm{\theta} \vert
\bm{\theta}') $$
where the quantities $Q$ and $H$ are defined as follows :
\begin{eqnarray*}
  Q(\bm{\theta} \vert \bm{\theta}')&=&\mathbb{E}\left[ \log \mathbb{P}_{\bm{\theta}} (\bm{Y},\bm{S}) \vert \bm{Y}=\bm{y}, \bm{\theta}'\right] \\
  H(\bm{\theta} \vert \bm{\theta}') &=&\mathbb{E} \left[ \log \mathbb{P}_{\bm{\theta}} (\bm{Y},\bm{S} \vert \bm{Y}=\bm{y}) \vert \bm{y}, \bm{\theta}' \right]
\end{eqnarray*}

The  EM algorithm is  divided in  two steps:  E-step (Expectation)  and M-step
(Maximization). Both steps consist  of, respectively, computing and maximizing
the   function   $Q(\bm{\theta}  \vert   \bm{\theta}^{(k)})$,   that  is   the
log-likelihood of the complete model  conditional on the observed sequence $y$
and  on the  current parameter  $\bm{\theta}^{(k)}$. Using  the fact  that the
function $\bm{\theta}  \rightarrow H(\bm{\theta} \vert  \bm{\theta}^{(k)})$ is
maximal in  $\bm{\theta}^{(k)}$, Dempster et  al.  proved that  this procedure
necessarily  increases   the  log-likelihood  $L_{\bm{y}}(\bm{\theta})$.   See
\cite{wu}  for  a detailed  study  of the  convergence  properties  of the  EM
algorithm.

We  now derive  analytical expressions  for both  E-step and  M-step.  In this
particular    case,    the     log-likelihood    of    the    complete    data
$(\bm{Y}_{m+1}^n,\bm{S}_{m+1}^n)$  conditional on  the first  $m$ observations
$\bm{Y}_1^m$ writes:
\begin{multline}\label{logl}
  \log \mathbb{P}_{\bm{\theta}}(\bm{Y}_{m+1}^n,\bm{S}_{m+1}^n \vert
  \bm{Y}_1^m)=\sum_{t=m+1}^n \sum_{g=1}^m \sum_{i \in \y} \sum_{j \in \y} 1\mskip-5mu\relax\mathrm{l}_{\lbrace Y_{t-g}=i, Y_t=j,S_t=g\rbrace} \log \pi_g(i,j)\\
  +\sum_{t=m+1}^n       \sum_{g=1}^m      1\mskip-5mu\relax\mathrm{l}_{\lbrace
    S_t=g\rbrace} \log \varphi_g.
\end{multline}

\paragraph{E-step}
The Estimation step is computing the expectation of this function (\ref{logl})
conditional  on   the  observed  data  $\bm{y}$  and   the  current  parameter
$\bm{\theta}^{(k)}$, that  is calculating, for  all $t>m$ and for  all element
$g$ in $\lbrace 1,...,m\rbrace$, the following quantity,
\begin{equation}\label{pRegime}
  \mathbb{E}(1\mskip-5mu\relax\mathrm{l}_{\lbrace S_t=g \rbrace} \vert \bm{y}, \bm{\theta}^{(k)})=\mathbb{P}(S_t=g \vert \bm{y},\bm{\theta}^{(k)}).
\end{equation}
Then, function Q writes:
\begin{multline}\label{Q}
  Q(\bm{\theta} \vert \bm{\theta}^{(k)})= \sum_{t=m+1}^n \sum_{g=1}^m \sum_{i
    \in \y} \sum_{j \in \y} \left [\mathbb{P}(S_t=g\vert \bm{y},\bm{\theta}^{(k)}) \log \pi_g(i,j) \right] 1\mskip-5mu\relax\mathrm{l}_{\lbrace y_{t-g}=i, y_t=j\rbrace}\\
  +    \sum_{t=m+1}^n     \sum_{g=1}^m    \mathbb{P}(S_t=g    \vert    \bm{y},
  \bm{\theta}^{(k)})\log \varphi_g.
\end{multline}

So E-step reduces to computing the probabilities (\ref{pRegime}), for which we
derive an explicit  expression by using the theory of  graphical models in the
particular  case  of  DAG  structured dependencies  \cite{lauritzen}.   First,
remark that  the state  variable $S_t$ depends  on the sequence  $\bm{Y}$ only
through the $m+1$ variables $\lbrace Y_{t-m},...,Y_{t-1},Y_t \rbrace$:
\begin{equation}\label{pgmot}
  \forall t \leq n, \forall g \in \lbrace 1,...,m\rbrace,\ \
  \mathbb{P}(S_t=g \vert \bm{y}, \bm{\theta})=\mathbb{P}(S_t=g \vert \bm{Y}_{t-m}^t=\bm{y}_{t-m}^t, \bm{\theta}).
\end{equation}

%%%%%%%%%%%%%%%%%%%%%%%%%%%%%%%%%%%%%%%%%%%%%%%%%%%%%%%%%%%%%%%%%%%%%%%%%%%%
\begin{figure}
\center
\includegraphics[width=0.7\textwidth]{\PathFig 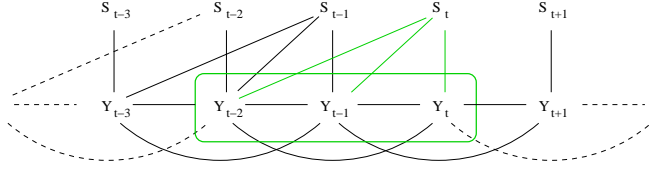}
\caption{Moral graph of a $2^{nd}$ order MTD$_1$ model.}
\label{fig:moralgraphe}
\end{figure}
%%%%%%%%%%%%%%%%%%%%%%%%%%%%%%%%%%%%%%%%%%%%%%%%%%%%%%%%%%%%%%%%%%%%%%%%%%%%

\noindent Indeed, independence properties can  be derived from the moral graph
(Fig.  \ref{fig:moralgraphe}) which is obtained  from the DAG structure of the
dependencies (Fig. \ref{fig:dependance}) by  ``marrying'' the parents, that is
adding an edge between the common  parents of each variable, and then deleting
directions.  In this moral  graph, the set $\lbrace Y_{t-m},...,Y_{t}\rbrace $
separates  the  variable  $S_t$  from   the  rest  of  the  sequence  $\lbrace
Y_{1},...,Y_{t-m-1}\rbrace   $   so   that   applying  corollary   3.23   from
\cite{lauritzen}  yields:  $$S_t  \indep  (\bm{Y}_1^{t-m-1},\bm{Y}_{t+1}^n)  \
\vert \ \bm{Y}_{t-m}^t$$

From  now on,  we denote  $\bm{i}_m^0=i_mi_{m-1}...i_1i_0$  any $(m+1)$-letter
word composed of  elements of $\mathcal{Y}$.  For all  $g$ in $\lbrace 1,...,m
\rbrace$,  for  all $\bm{i}_m^0$  elements  of  $\mathcal{Y}$, Bayes'  Theorem
gives:

\begin{eqnarray}\label{pgmot2}
  \mathbb{P}(&S_t&=g\vert Y_{t-m}^{t}=\bm{i}_m^0, \bm{\theta})\nonumber\\
  &=&\frac{\mathbb{P}(S_t=g,Y_t=i_0 \vert \bm{Y}_{t-m}^{t-1}=\bm{i}_m^1,\bm{\theta})}{\mathbb{P}(Y_t=i_0 \vert \bm{Y}_{t-m}^{t-1}=\bm{i}_m^1,\bm{\theta})}\nonumber\\
  &=&\frac{ \mathbb{P}(Y_t=i_0 \vert S_t=g,\bm{Y}_{t-m}^{t-1}=\bm{i}_m^1,\bm{\theta}) \mathbb{P}(S_t=g \vert \bm{Y}_{t-m}^{t-1}=\bm{i}_m^1,\bm{\theta})}{\sum_{l=1}^m   \mathbb{P}(Y_t=i_0 \vert S_t=l,\bm{Y}_{t-m}^{t-1}=\bm{i}_m^1,\bm{\theta})\mathbb{P}(S_t=l\vert \bm{Y}_{t-m}^{t-1}=\bm{i}_m^1,\bm{\theta})}.
\end{eqnarray}

\noindent  We  show below  that  the  probabilities $\mathbb{P}(Y_t=i_0  \vert
S_t=g,\bm{Y}_{t-m}^{t-1}=\bm{i}_m^1,\bm{\theta})$  and $\mathbb{P}(S_t=g \vert
\bm{Y}_{t-m}^{t-1}=\bm{i}_m^1,\bm{\theta})$  in expression  (\ref{pgmot2}) are
entirely explicit.   First, conditional on $\bm{\theta}$, the  state $S_t$ and
the variables $\bm{Y}_{t-m}^{t-1}$, the distribution of $Y_t$ writes:

$$\mathbb{P}(Y_t=i_0 \vert S_t=g,\bm{Y}_{t-m}^{t-1}=\bm{i}_m^1,\bm{\theta})=\bm{\pi}_g(i_g,i_0).$$

\begin{figure}
  \center \includegraphics[width=0.7\textwidth]{\PathFig 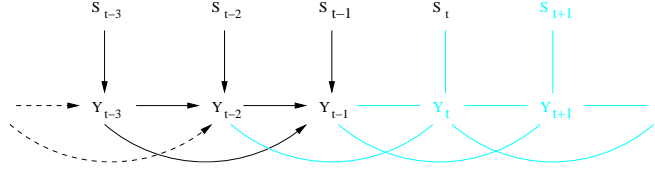}
  \caption{In black: graph of the  smallest ancestral set containing $S_t$ and
    the two variables $(Y_{t-2},Y_{t-1})$ in the particular case of a $2^{nd}$
    order MTD$_1$  model.  (The part of  the structure dependency  DAG that is
    excluded from the smallest ancestral set appears here in light blue.)}
\label{fig:DAGanc}
\end{figure}

\begin{figure}
\hspace{1.7cm}
\includegraphics[width=0.5\textwidth]{\PathFig 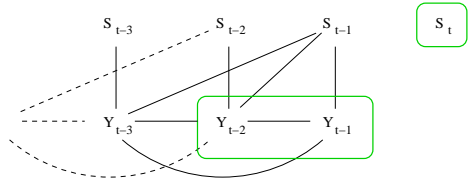}
\caption{Moral graph of the smallest ancestral set in Figure \ref{fig:DAGanc}. There is no path between
  $S_t$ and the subset of $2$ variables $\{Y_{t-2},Y_{t-1}\}$.}
\label{fig:moralDAGanc}
\end{figure}

\noindent Second, although the state  $S_t$ depends on the $(m+1)$-letter word
$\bm{Y}_{t-m}^t$, which brings information about the probability of transition
from $\bm{Y}_{t-m}^{t-1}$ to $Y_t$, it  does not depend on the $m$-letter word
formed by  the only variables  $\bm{Y}_{t-m}^{t-1}$.  This again  follows from
the  same corollary in  \cite{lauritzen}.  The  independence of  the variables
$S_t$  and $\bm{Y}^{t-1}_{t-m}$  is derived  from  the graph  of the  smallest
ancestral  set containing  these variables,  that is  the  subgraph containing
$S_t$, $\bm{Y}^{t-m}_{t-1}$ and the whole  line of their ancestors (See Figure
\ref{fig:DAGanc} for  an illustration  when $n=2$).  It  turns out  that, when
considering the moralization  of this subgraph (Figure \ref{fig:moralDAGanc}),
there  is  no path  between  $S_t$  and  the set  $\bm{Y}_{t-m}^{t-1}$.   This
establishes  $S_t \indep  \bm{Y}_{t-m}^{t-1}$ and  we  have $$\mathbb{P}(S_t=g
\vert     \bm{Y}_{t-m}^{t-1}=\bm{i}_m^1,\bm{\theta})=\mathbb{P}(S_t=g    \vert
\bm{\theta})=\varphi_g.$$

%%%%%%%%%%%%%%%%%%%%%%%%%%%%%%%%%%%%%%%%%%%%%%%%%%%%%%%%%%%%%%%%%%%%%%%%%%%%
%%%%%%%%%%%%%%%%%%%%%%%%%%%%%%%%%%%%%%%%%%%%%%%%%%%%%%%%%%%%%%%%%%%%%%%%%%%%
%%%%%%%%%%%%%%%%%%%%%%%%%%%%%%%%%%%%%%%%%%%%%%%%%%%%%%%%%%%%%%%%%%%%%%%%%%%%

\noindent Finally,  the probability (\ref{pgmot2}), is  entirely determined by
the current parameter $\bm{\theta}$ and does not depend on the time $t$.

As  a   result,  the  $k^{th}$   iteration  of  Estimation-step   consists  in
calculating, for all $g$ in $ \lbrace1,...,m\rbrace$ and for all $m+1$-letters
word $\bm{i}_m^0$ of elements of $\mathcal{Y}$,
\begin{multline}\label{Estep}
\forall g \in \lbrace1,...,m\rbrace, \forall \ i_m, ..., i_1,i_0 \in \y,\\
\mathbb{P}_S^{(k)}(g \vert \bm{i}_m^0) 
=\mathbb{P}(S_t=g \vert \bm{Y}_{t-m}^t=\bm{i}_m^0,\bm{\theta}^{(k)})
= \frac {\varphi_g^{(k)} \pi_g^{(k)}(i_g,i_0)}{\sum_{l=1}^m \varphi_l^{(k)} \pi_l^{(k)}(i_l,i_0) }. 
\end{multline}

\paragraph{M-Step}
Maximization  of the  function $Q(\bm{\theta}  \vert  \bm{\theta}^{(k)})$ with
respect to  the constraints  imposed on the  vector $\bm{\varphi}$ and  on the
elements  of  the transition  matrices  $\bm{\pi}_1,...,\bm{\pi}_m$ is  easily
achieved using Lagrange method:\break  $\forall g \in \lbrace 1,...,m \rbrace,
\forall i,j \in \y$,

\begin{eqnarray}
\varphi_g^{(k+1)}\!\!&=&\!\!\frac{1}{n-m}\sum_{i_m...i_0}\mathbb{P}^{(k)}(g\vert \bm{i}_m^0) N(\bm{i}_m^0)\label{MstepPhi}\\
\bm{\pi}_g^{(k+1)}(i,j)\!\!&=&\!\!\frac{\sum_{i_m...i_{g+1}i_{g-1}...i_1}\mathbb{P}^{(k)}(g\vert
  \bm{i}_m^{g+1}i\bm{i}_{g-1}^1j)
  N(\bm{i}_m^{g+1}i\bm{i}_{g-1}^1j)}{\sum_{i_m...i_{g+1}i_{g-1}...i_1i_0}
  \mathbb{P}^{(k)}(g\vert
  \bm{i}_m^{g+1}i\bm{i}_{g-1}^0)N(\bm{i}_m^{g+1}i\bm{i}_{g-1}^0)}\label{MstepPi} 
\end{eqnarray}

\noindent  where   sums  are  carried   out  for  the  variables   $i_m,  ...,
i_{g+1},i_{g-1}, ...,  i_1, i_0$ taking values  in $\y$, $n$ is  the length of
the observed  sequence $\bm{y}$ and $N(\bm{i}_m^0)$ the  number of occurrences
of the word $\bm{i}_m^0$ in this sequence.

\paragraph{Initialization}
To maximize  the chance of reaching  the global maximum, we  run the algorithm
from various  starting points. One initialization is  derived from contingency
tables  between  each lag  $y_{t-g}$  and the  present  $y_t$  as proposed  by
Berchtold  \cite{berchtold} and  several others  are randomly  drawn  from the
uniform distribution.

\paragraph{EM-Algorithm for MTD models}
\textit{
\begin{itemize}
\item Compute the number of occurrences of each $(m+1)$-letters word $N(\bm{i}_m^0)$,
\item initialize parameters $(\bm{\varphi}^{(0)}, \bm{\pi}^{(0)})$,
\item choose a  stopping rule, \textit{i.e.} an upper  threshold $\varepsilon$ on the
  increase of the log-likelihood, 
\item     iterate     E     and     M     steps     given     by     equations
  (\ref{Estep},\ref{MstepPhi},\ref{MstepPi}),
\item stop when $ L_y( \bm{\theta}^{(k+1)})-L_y( \bm{\theta}^{(k)}) < \varepsilon$.
\end{itemize}}

A software implementation of our algorithm is available in the library seq$++$ at
\texttt{http://stat.genopole.cnrs.fr/seqpp}.

\section{Applications}\label{section:appli}

\subsection{Comparison with Berchtold's Estimation}\label{subsection:appliBerchtold}
%%%%%%%%%%%%%%%%%%%%%%%%%%%%%%%%%%%%%%%%%%%%%%%%%%%%%%%%%%%%%%%%%%%%%%%%%%%%%%
\begin{table}
\begin{center}
\caption{Maximum  log-likelihood  of  MTD$_1$   models  estimated  by  EM  and
  Berchtold's algorithm (see \cite{berchtold}, section 5.1 and 6.2).} 
\label{tab:logL}
\begin{tabular}{|c|c|c|c|c|}
  \hline
  Order $m$ &$q=\vert \mathcal{Y} \vert$ &Berchtold &EM &Sequence\\
  \hline
  \hline
  2&3&-486.4&-481.8&Pewee\\
  &4&-1720.1 &-1718.5 &$\alpha$A-Crystallin\\
  \hline
  3&3&-484.0&-480.0&Pewee\\
  &4&-1710.6&-1707.9&$\alpha$A-Crystallin\\
  \hline
\end{tabular}
\end{center}
\end{table}

In this  paper, we focus  on estimation of  the MTD$_l$ model  (see Definition
\ref{def:MTDl}) which has a specific but same order matrix transition for each
lag. We  evaluate the performance of  the EM algorithm with  comparison to the
last  and best  algorithm to  date, developed  by  Berchtold \cite{berchtold}. 
Among  others, Berchtold  estimates the  parameters of  MTD$_l$ models  on two
sequences analyzed in previous articles: a  time serie of the twilight song of
the wood pewee and the  mouse $\alpha$A-Crystallin Gene sequence (the complete
sequences appear in \cite{rt94}, Tables 7 and 12).  The song of the wood pewee
is a sequence composed of 3 distinct phrases (referred to as $1,2,3$), whereas
the $\alpha$A-Crystallin Gene is composed of 4 nucleotides: a, c, g, t.

We apply  our estimation  method to these  sequences and obtain  comparable or
higher value of the log-likelihood for both (see Tab.  \ref{tab:logL}).  Since
the original parametrization of the MTD$_1$  model is not injective, it is not
reasonable to compare their values.  To overcome this problem, we computed the
parameters  from the  set  $\bar{\Theta}_u$ defined  in (\ref{theta_u}).   The
estimated parameters  (using a precision  parameter $\varepsilon =  0.001$) of
the $2^{nd}$ order MTD$_1$ model on the  song of wood Pewee (first line of the
Table  \ref{tab:logL}) are exposed  in Figure  \ref{fig:EstimPewee}.  Complete
results appear  in appendix \ref{appendix:estim},  namely estimated parameters
$\hat{\bm{\varphi}},    \hat{\bm{\pi}}_1,    \hat{\bm{\pi}}_2$    and    their
corresponding full $2^{nd}$ order transition matrices $\hat{\bm{\Pi}}$.

%%%%%%%%%%%%%%%%%%%%%%%%%%%%%%%%%%%%%%%%%%%%%%%%%%%%%%%%%%%%%%%%%%%%%%%%%%%%%%
\begin{figure}
\caption{Estimation of a $2^{nd}$ order MTD$_1$ model on the song of the wood
pewee.  We use u=1 (song n°1) as reference letter to express the parameters
 defined in (\ref{theta_u}).}
\label{fig:EstimPewee}
\paragraph{}
Estimates obtained with:
\begin{itemize}
\item Berchtold's algorithm  ($L_y(\hat{\bm{\theta}})=-486.4$):
\[
\begin{array}{cc}
\left[\hat{p}_1(1;i,j)\right]_{1 \leq i,j \leq 3}= &
\left(\begin{array}{ccc}
0.754169   & 0.198791   & 0.073356\\
0.991696   & 0.         & 0.03462 \\
0.993579   & 0.003497   & 0.02924 \end{array}
\right)
\end{array}
\]

\[
\begin{array}{cc}
\left[\hat{p}_1(2;i,j)\right]_{1 \leq i,j \leq 3}= &
\left(\begin{array}{ccc}
0.754169   & 0.198791   & 0.073356\\
0.137205   & 0.213411   & 0.649384\\
0.048023   & 0.927598   & 0.044116\end{array}
\right)
\end{array}
\]
\vspace{0.2cm}

\item EM-algorithm ($L_y(\hat{\bm{\theta}})=-481.8$):
\[
\begin{array}{cc}
\left[\hat{p}_1(1;i,j)\right]_{1 \leq i,j \leq 3}=&
\left(\begin{array}{ccc}
0.75305   &  0.200475  &  0.046475\\
0.991475  &  0.        &  0.008525\\
0.996425  &  0.003575  &  0.  \end{array}
\right)
\end{array}
\]

\[
\begin{array}{cc}
\left[\hat{p}_1(2;i,j)\right]_{1 \leq i,j \leq 3}=  &
\left(\begin{array}{ccc}
0.75305  &   0.200475  &  0.046475\\
0.137525 &   0.21135   &  0.651125\\
0.02805  &   0.925475  &  0.046475\end{array}
\right)
\end{array}
\]
\end{itemize}
\end{figure}
%%%%%%%%%%%%%%%%%%%%%%%%%%%%%%%%%%%%%%%%%%%%%%%%%%%%%%%%%%%%%%%%%%%%%%%%%%%%%%

For  both  sequences  under  study,  Pewee and  $\alpha$A-crystallin,  EM  and
Berchtold algorithms lead to  comparable estimations.  The EM algorithm proves
here to be  an effective method to maximize the  log-likelihood of MTD models. 
Nevertheless,  EM algorithm  offers  the advantage  to  be very  easy to  use. 
Whereas Berchtold's algorithm requires to  set and update a parameter $\delta$
to alter the vector $\bm{\varphi}$  and each row of the matrices $\bm{\pi}_g$,
running  the  EM   algorithm  only  requires  the  choice   of  the  threshold
$\varepsilon$ in the stopping rule.

\subsection{Estimation on DNA coding sequences}\label{subsection:estimDNA}

DNA coding  regions are translated into  proteins with respect  to the genetic
code, which is defined on  blocks of three nucleotides called \textit{codons}. 
Hence,  the nucleotides  in these  regions are  constrained in  different ways
according to  their position in the  codon. It is common  in bioinformatics to
use  three different  transition matrices  to predict  the nucleotides  in the
three positions of the codons. This model is called the \textit{phased} Markov
model.

Since  we aim  at  comparing  the goodness-of-fits  of  models with  different
dimensions, the maximal  value of a penalized likelihood  function against the
dimension of parameter  space will be used to assess  each model. The Bayesian
Information criterion  \cite{schwarz78} for this evaluation is  defined as: $$
BIC(\mathcal{M})  = -2  L_y(\hat{\bm{\theta}}_{\mathcal{M}})  + d(\mathcal{M})
\log  n , $$
where $\hat{\bm{\theta}}_{\mathcal{M}}$  stands for  the maximum
likelihood  estimate  of model  $\mathcal{M}$.   The  lower  the BIC  a  model
achieves, the more pertinent it is.

\begin{figure}
\begin{center}
\includegraphics[width=0.7\textwidth]{\PathFig 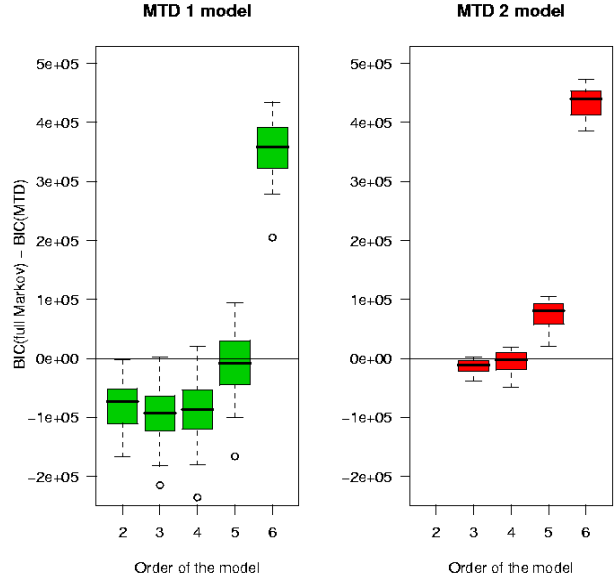}
\end{center}
\caption{Difference according to the BIC  criterion between MTD models and the
  corresponding fully parametrized Markov Model.}
   \label{fig:diffBIC}
\end{figure}

BIC evaluation has been computed on DNA coding sequence sets from bacterial
genomes. Each of  these sequence sets has  length ranging from 1 500  000 to 5
000 000. Displayed values in Figure \ref{fig:diffBIC} are
averages over  the 15 sequences  set of the  difference between the  BIC value
achieved by the full Markov model and the one achieved by the MTD model of the
same  order.  Whenever this  figure  is  positive, the  MTD  model  has to  be
preferred to the full Markov model.

%%%%%%%%%%%%%%%%%%%%%%%%%%%%%%%%%%%%%%%%%%%%%%%%%%%%%%%%%%%%%%%%%%%%%%%%%%%%%

The full Markov model turns out to outperform the MTD$_1$ model when the order
is inferior to 4. This is not surprising since the estimation is computed over
large datasets that provide a sufficient amount of information with respect to
the  number of  parameters  of the  full  model. However,  the 5$^{th}$  order
MTD$_1$  model and  full Markov  model have  comparable performances,  and the
MTD$_1$ model outperforms the full Markov  model for higher orders. This is an
evidence that although MTD$_1$ only  approximate the full Markov models, their
estimation accuracy decreases slower with the order.

Even more striking is the comparison of the MTD$_2$ model with the full Markov
model.  Whatever the  order  of the  model,  its goodness-of-fit  is at  least
equivalent to  the one achieved  by the full  Markov model. The  MTD$_l$ model
turns  out to  be a  satisfactory trade-off  between dimension  and estimation
accuracy.

\section{Acknowledgments}
We  thank  Bernard Prum  and  Catherine  Matias  for their  very  constructive
suggestions, and Vincent  Miele for his implementation of  the EM algorithm in
the seq++  library.  Moreover,  we thank the  referees for their  comments and
suggestions which improve this paper.

\appendix
\normalsize
\begin{singlespace}
  \section{Example  of   equivalent  parameters  defining   the  same  MTD$_1$
    model}\label{appendix:2eq}
Let  the size  state  space be  4  as for  DNA sequences  $\mathcal{Y}=\lbrace
a,c,g,t  \rbrace$  and  consider   these  two  $2^{nd}$  order  MTD$_1$  model
parameters $\bm{\theta}, \bm{\theta}'$.

%\begin{multicols}{3}{

\[
\begin{array}{lccc}

\bm{\varphi} = (0.3 , 0.7) &

\bm{\pi}_1= 
\left(\begin{array}{cccc}
 0.1  &  0.2  &  0.3  &  0.4 \\
 0.4  &  0.3  &  0.2  &  0.1 \\
 0.2  &  0.2  &  0.2  &  0.4 \\
 0.4  &  0.2   & 0.2  &  0.2 \end{array}
\right)&

\bm{\pi}_2= 
\left(\begin{array}{cccc}
0.1  &  0.1 &   0.1  &  0.7 \\
0.2  &  0.2 &   0.4  &  0.2 \\
0.3  &  0.3 &   0.3  &  0.1 \\
0.3  &  0.2 &   0.3  &  0.2 \end{array}
\right)\\
&&\\

\bm{\varphi}' = (0.2, 0.8)&

\bm{\pi}_1'= 
\left(\begin{array}{cccc}
0.2  &   0.1   &  0.2   &  0.5  \\
0.65 &   0.25  &  0.05  &  0.05 \\
0.35 &   0.1   &  0.05  &  0.5  \\
0.65 &   0.1   &  0.05  &  0.2 \end{array}
\right)&

\bm{\pi}_2'= 
\left(\begin{array}{cccc}
0.075  &   0.1375 &   0.15   &   0.6375 \\
0.1625 &   0.225  &   0.4125 &   0.2    \\
0.25   &   0.3125 &   0.325  &   0.1125 \\
0.25   &   0.225  &   0.325  &   0.2   \end{array}
\right)%%

\end{array}
\]

\noindent Both parameters define the same $2^{nd}$ order Markov transition
matrix $\bm{\Pi}$. 

\[
\begin{array}{cc}
&  a \hspace{0.7cm}    c \hspace{0.7cm}   g \hspace{0.7cm}   t \\
\bm{\Pi}= 
\begin{array}{l}
aa\\
ac\\
ag\\
at\\
ca\\
cc\\
cg\\
ct\\
ga\\
gc\\
gg\\
gt\\
ta\\
tc\\
tg\\
tt\end{array}
&
\left(
\begin{array}{cccc}
0.1  &   0.13  &  0.16 &   0.61 \\
0.19 &   0.16  &  0.13 &   0.52 \\
0.13 &   0.13  &  0.13 &   0.61 \\
0.19 &   0.13  &  0.13 &   0.55 \\
0.17 &   0.2   &  0.37 &   0.26 \\
0.26 &   0.23  &  0.34 &   0.17 \\
0.2  &   0.2   &  0.34 &   0.26 \\
0.26 &   0.2   &  0.34 &   0.2  \\
0.24 &   0.27  &  0.3  &   0.19 \\
0.33 &   0.3   &  0.27 &   0.1  \\
0.27 &   0.27  &  0.27 &   0.19 \\
0.33 &   0.27  &  0.27 &   0.13 \\
0.24 &   0.2   &  0.3  &   0.26 \\
0.33 &   0.23  &  0.27 &   0.17 \\
0.27 &   0.2   &  0.27 &   0.26 \\
0.33 &   0.2   &  0.27 &   0.2 \end{array}
\right)
\end{array}
\]

\section{EM algorithm for other MTD models}\label{appendix:mtdl}

\subsection{Single matrix MTD model: iteration k.}

\paragraph{E-Step}

$\forall g \in \lbrace1,...,m\rbrace, \forall  \ i_m, ..., i_1,i_0 \in \lbrace
1,...,q \rbrace,$
$$\mathbb{P}_S^{(k)}(g \vert i_m^0) = \frac {\varphi_g^{(k)} \bm{\pi}^{(k)}(i_g,i_0)}{\sum_{l=1}^m \varphi_l^{(k)} \bm{\pi}^{(k)}(i_l,i_0) }. $$

\paragraph{M-Step}
$\forall    g    \in    \lbrace    1,...,m   \rbrace,    \forall    i,j    \in
\lbrace1,...,q\rbrace$,

\begin{eqnarray*}
  \varphi_g^{(k+1)}\!\!&=&\!\!\frac{1}{n-m}\sum_{i_m...i_0}\mathbb{P}^{(k)}(g\vert \bm{i}_m^0) N(\bm{i}_m^0)\\
  \bm{\pi}^{(k+1)}(i,j)\!\!&=&\!\!\frac{\sum_{g=1}^m\sum_{i_m...i_{g+1}i_{g-1}...i_1}\mathbb{P}^{(k)}(g\vert
    \bm{i}_m^{g+1}i\bm{i}_{g-1}^1j)
    N(\bm{i}_m^{g+1}i\bm{i}_{g-1}^1j)}{\sum_{g=1}^m\sum_{i_m...i_{g+1}i_{g-1}...i_1i_0}
    \mathbb{P}^{(k)}(g\vert
    \bm{i}_m^{g+1}i\bm{i}_{g-1}^0)N(\bm{i}_m^{g+1}i\bm{i}_{g-1}^0)}
\end{eqnarray*}

\noindent  \textit{where sums  are carried  out for  the variables  $i_m, ...,
  i_{g+1},i_{g-1}, ..., i_1,  i_0$ varying from $1$ to $q$,  $n$ is the length
  of the observed  sequence $y$ and $N(\bm{i}_m^0)$ the  number of occurrences
  of the word $\bm{i}_m^0$ in this sequence.}

\subsection{MTD$_l$ model: iteration k.}
\paragraph{E-Step}
%\begin{multline*}
 $ \forall g \in \lbrace 1,...,m-l+1\rbrace, \forall i_m,...i_1,i_0 \in \lbrace 1,...,q \rbrace,$
$$  \mathbb{P}_S^{(k)}(g      \vert      \bm{i}_m^0)=     \frac      {\varphi_g^{(k)}
    \bm{\pi}_g^{(k)}(\bm{i}_{g+l-1}^g,i_0)}{\sum_{h=1}^{m-l+1}            \varphi_h^{(k)}
    \bm{\pi}_h^{(k)}(\bm{i}_{h+l-1}^h,i_0) }.$$
%\end{multline*}

\paragraph{M-Step}
$\forall   g  \in   \lbrace   1,...,m  \rbrace,   \forall  i_l,...,i_1,j   \in
\lbrace1,...,q\rbrace$,
\begin{eqnarray*}
  \varphi_g^{(k+1)}\!\!&=&\!\!\frac{1}{n-m}\sum_{u_m...u_0}\mathbb{P}_S^{(k)}(g\vert \bm{u}_m^0) N(\bm{u}_m^0)\\
  \bm{\pi}_g^{(k+1)}(i_li_{l-1}...i_1,j)\!\!&=&\!\!\frac{\sum_{u_m...u_{g+l}u_{g-1}...u_1}\mathbb{P}_S^{(k)}(g\vert \bm{u}_m^{g+l}\bm{i}_l^1\bm{u}_{g-1}^1j) 
    N(\bm{u}_m^{g+l}\bm{i}_l^1\bm{u}_{g-1}^1j)}{\sum_{u_m...u_{g+l}u_{g-1}...u_1u_0}
    \mathbb{P}_S(g\vert \bm{u}_m^{g+l}\bm{i}_l^1\bm{u}_{g-1}^0) N(\bm{u}_m^{g+l}\bm{i}_l^1\bm{u}_{g-1}^0)},
\end{eqnarray*}

\noindent \textit{ where  sums are  carried out for  the variables $u_m,  ..., u_{g+l},
u_{g-1}, ...,  u_1, u_0$ varying  from $1$  to $q$, $n$  is the length  of the
observed sequence  $y$ and  $N(\bm{i}_m^0)$ the number  of occurrences of  the word
$\bm{i}_m^0$ in this sequence.}

\section{$2^{nd}$ order MTD$_1$ estimates obtained on both the 
 song of wood pewee and the  mouse $\alpha$A-Crystallin Gene sequence (Section
 \ref{subsection:appliBerchtold}).}\label{appendix:estim}

\begin{enumerate}
\item \textbf{Song of wood pewee}

Berchtold's algorithm (see \cite{berchtold}, section 5.1): $L_y(\hat{\bm{\theta}})=-486.4$. 

\[
\begin{array}{lccc}
\hat{\bm{\varphi}} = (0.269,0.731)&
\hat{\bm{\pi}}_1= 
\left(\begin{array}{ccc}
0.097 & 0.739 & 0.164 \\
0.980 & 0 & 0.020 \\
0.987 & 0.013 & 0\end{array}
\right)&
\hat{\bm{\pi}}_2= 
\left(\begin{array}{ccc}
0.996 & 0 & 0.004\\
0.152 & 0.020 & 0.828\\
0.003 & 0.997 & 0 \end{array}
\right).\\
\end{array}
\]

\vspace{0.3cm}
\noindent 
EM-algorithm: $L_y(\hat{\bm{\theta}})=-481.8)$.

\[
\begin{array}{lccc}

\hat{\bm{\varphi}} = (0.275,0.725)&
\hat{\bm{\pi}}_1= 
\left(\begin{array}{ccc}
0.102 & 0.729 & 0.169\\
0.969 & 0 & 0.031\\
0.987 & 0.013  &  0\end{array}
\right)&
\hat{\bm{\pi}}_2= 
\left(\begin{array}{ccc}
1 &  0 &  0 \\
0.151 &  0.015 &  0.834 \\
0 &  1 &  0 \end{array}
\right).\\
\end{array}
\]

\noindent  These  estimated   parameters  define  respectively  the  following
$2^{nd}$ order Markov transition matrices $\hat{\bm{\Pi}}_B$ and $\hat{\bm{\Pi}}_{EM}$. 

\begin{small}
\[
\begin{array}{cc}

\hat{\bm{\Pi}}_B= \!\! 
\left(\begin{array}{lll}
0.754169 &    0.198791 &   0.047040\\
0.991696 &    0.       &   0.008304\\
0.993579 &    0.003497 &   0.02924 \\
0.137205 &    0.213411 &   0.649384\\
0.374732 &    0.01462  &   0.610648\\
0.376615 &    0.018117 &   0.605268\\
0.028286 &    0.927598 &   0.044116\\
0.265813 &    0.728807 &   0.00538 \\
0.267696 &    0.732304 &   0.      \end{array}
\right)
 &
\hat{\bm{\Pi}}_{EM}\!= \!\!
\left(\begin{array}{lll}
0.75305  &   0.200475  &  0.046475\\
0.991475 &   0.        &  0.008525\\
0.996425 &   0.003575  &  0.      \\
0.137525 &   0.21135   & 0.651125\\
0.37595  &   0.010875  &  0.613175\\
0.3809   &   0.01445   &  0.60465 \\
0.02805  &   0.925475  &  0.046475\\
0.266475 &   0.725     &  0.008525\\
0.271425 &   0.728575  &  0.      \end{array}
\right)

\end{array}
\]

\end{small}
\newpage
\item \textbf{Mouse $\alpha$A-Crystallin Gene sequence}

\noindent %\textbf{
EM-algorithm: $L_y(\hat{\bm{\theta}})=-1718.5$.
$$\hat{\bm{\varphi}} = (0.562,0.438),$$
\[
\begin{array}{ccc}

\hat{\bm{\pi}}_1= 
\left(\begin{array}{cccc}
0.225 & 0.140 & 0.506 & 0.129\\
0.354& 0.300 & 0.008 & 0.338 \\
0.271& 0.123 & 0.456& 0.150 \\
0.166& 0.191& 0.430 & 0.213\end{array}
\right)&
\hat{\bm{\pi}}_2= 
\left(\begin{array}{cccc}
0.094 & 0.600 & 0.149 & 0.157\\
0.335& 0.271 & 0.153 & 0.241\\
0.185 & 0.415 & 0.099 & 0.301\\
0.192& 0.370 & 0.129& 0.309  \end{array}
\right).\\
\end{array}
\]

\noindent  These  estimated   parameters  define  respectively  the  following
$2^{nd}$ order Markov transition matrix $\hat{\bm{\Pi}}_{EM}$. 

\begin{small}
\[
\begin{array}{c}

\hat{\bm{\Pi}}_{EM}\!= \!\!
\left(\begin{array}{llll}
0.167622 &0.341480 &0.349634 &0.141264\\
0.240120 &0.431400 &0.069758 & 0.258722\\
0.193474 &0.331926 &0.321534 &0.153066\\
0.134464 &0.370142 &0.306922 &0.188472\\
0.273180 &0.197378 &0.351386 &0.178056\\
0.345678 &0.287298 &0.071510 &0.295514\\
0.299032 &0.187824 &0.323286 &0.189858\\
0.240022 &0.226040 &0.308674 &0.225264\\
0.207480 &0.260450 &0.327734 &0.204336\\
0.279978 &0.350370 &0.047858 &0.321794\\
0.233332 &0.250896 &0.299634 &0.216138\\
0.174322 &0.289112 &0.285022 &0.251544\\
0.210546 &0.240740 &0.340874 &0.207840\\
0.283044 &0.330660 &0.060998 &0.325298\\
0.236398 &0.231186 &0.312774 &0.219642\\
0.177388 &0.269402 &0.298162 &0.255048     \end{array}
\right)

\end{array}
\]
\end{small}
No detail on the $2^{nd}$ order MTD$_1$ estimates from the mouse $\alpha$A-Crystallin Gene sequence
is given in \cite{berchtold}.

\end{enumerate}

\end{singlespace}

\bibliographystyle{natbib}
\bibliography{biblioMTD.bib}

\end{document}